# The morphology and dynamics of polymerization-induced phase separation

Kaifu Luo[*]

*Institut für Makromolekulare Chemie and Freiburger Materialforschungszentrum, Universität Freiburg, Stefan-Meier-Str. 31, D-79104 Freiburg, Germany and Laboratory of Physics, Helsinki University of Technology, P.O. Box 1100, FIN-02015 HUT, Espoo, Finland.*

**Abstract** The morphology and dynamics of polymerization-induced phase separation in the initially homogeneous solution of a non-reactive component in reactive monomers are investigated by incorporating the reaction kinetics into the time-dependent Ginzburg Landau equation. Analytical results show that there is a reduction of the initial length scale in the early stage of phase separation. The reason is the increase of the molecule weight of emerging polymer, independent of the fact whether the system goes through the metastable region or not. The numerical results are in good agreement with theoretical prediction quite well.[*]

---

[*] Author to whom correspondence should be addressed. E-mail: luokaifu@yahoo.com

## I. Introduction

When polymer blends are thermally quenched from the homogenous to the unstable region in phase diagram, phase separation occurs. Phase separation dynamics has been intensively investigated in the past years. It is clear that the dynamics of phase separation is controlled by concentration fluctuations in the early stage and by diffusion and interfacial tension in the late stage, respectively. The growth of the domain size in the late stage obeys the simple scaling, i.e. $R(t) \sim t^{\alpha}$, where α is the growth exponent and depends on the space dimension, hydrodynamic effects and composition.[1-6]

However, when two nonequilibrium phenomena proceed in a system at the same time, if one of them accompanies pattern formation the resulting pattern formation could be strongly affected by the competition between the two phenomena. The coupling between nonequilibrium phenomena is very interesting from both the scientific and industrial viewpoints because it may cause a new pattern formation. An example of such couplings is the competition between phase separation and chemical reaction. For polymer blends, studies on phase separation of a binary mixture A/B accompanied by the reversible reaction A⇔B were initiated by Glotzer et al..[7,8] Their theoretical and numerical results show that chemical reaction may suppress the phase separation and confine the phase-separated domains to microscopic length scale. Experimentally, fascinating patterns are observed in phase separation of binary polymer blends under photo-cross-linking reaction.[9–12]

For the initially homogeneous solution of a non-reactive component in reactive monomers, phase separation will occur in the course of polymerization. This process is known as polymerization-induced phase separation (PIPS).[13,14] PIPS has been widely investigated in recent years due to industrial importance.[15-32] The non-reactive component may be polymer or liquid crystal, such as the rubber modified epoxy and the polymer-dispersed liquid crystal (PDLC). For PIPS, instability of the system is driven by the progressive increase of the average molecular weight of the polymerizing species. Once the polymerization is initiated, the nonequilibrium structure will occur in a manner dependent on the competition between phase separation dynamics and reaction kinetics.

Experimentally, Kyu et al.[22] have studied phase separation in a polymerizing system consisting of carboxyl terminated polybutadiene acrylonitrile/epoxy/methylene dianiline, by means of time-resolved light scattering. An interesting phenomenon is the reduction of the initial length scale for the early period of the reaction. Theoretically, by incorporating polymerization kinetics into the Cahn-Hilliard equation, they simply analysed the dynamics of phase separation and attributed the reduction of the initial length scale for off-critical composition to nucleation initiated spinodal decomposition (NISD). In later experiments, the reduction of initial length scale was considered as criterion whether the phase separation goes through the nucleation growth (NG) mechanism or not.[16]

In this paper we show that the NISD is not the unique mechanism for the reduction of initial length scale. Thus it is not correct to consider the reduction of initial length scale as evidence that phase separation goes through the NG mechanism. The paper is organized as follows. The theoretical analysis is introduced in section II. Section III presents numerical results. Finally, Section IV gives conclusions.

**II. Theory**

In order to study the effect of the initial monomer concentration on the characteristic phase separation time, Lin and Taylor theoretically examined the PIPS process for PDLC by decoupling the phase separation and polymerization process based on the assumption that phase separation takes place at a well-defined instantaneous degree of polymerisation.[26] The model includes the Flory-Huggins free energy to describe phase separation and a kinetic equation for polymeriztion. To evaluate the phase separating process in PIPS, Ohnaga et al.[27] numerically investigated the time dependent concentration fluctuation using the non-linear Cahn-Hillard equation under the successive increase of quench depth. The model can describe qualitatively the morphological evolution during PIPS process for several epoxy-polymer mixtures. Similarly, Zhu[24] investigated the dynamics of PIPS by thermally induced phase separation with proper quench scheme using Monte Carlo simulation method. The results show that the phase separated domain size can be controlled by reaction rate and temperature. Recently, Kyu et al.[22] have incorporated the reaction kinetics into the time dependent Cahn-Hillard equation for describing the spatio-temporal growth of phase separated domains in a rubber modified epoxy subjected to PIPS. The reduction of the length scale was ascribed to nucleation. Later, Rey et al.[28-30] employed a simpler version of the same equation in the simulation of the morphology evolution in PIPS based on an idealized assumption of reaction-driven phase separation.

The numerical results show that both the dynamical and morphology in PIPS are sensitive to the magnitudes of the diffusion coefficient and reaction rate constant.

Here let's consider a mixture consisting of a non-reactive polymer A and a bifunctional reactive monomer B that undergoes linear polycondensation at fixed temperature. Following Kyu[22] and Rey[28-30], the total free energy functional, $F$, may be expressed as

$$\frac{F}{k_B T} = \int d\mathbf{r} \left[ \frac{f(\phi)}{k_B T} + \frac{a^2}{36\phi(1-\phi)} |\nabla\phi|^2 \right], \quad (1)$$

where $\phi$ is the order parameter, i.e., the volume fraction of the non-reactive polymer A, $a$ is the Kuhn's statistical segmental length and the same segmental length for monomer B and polymer A is assumed, $k_B$ is the Boltzmann constant, $T$ is the absolute temperature. The local free energy is generally described in term of Flory-Huggins theory, i.e.

$$\frac{f(\phi)}{k_B T} = \frac{\phi}{N_A}\ln\phi + \frac{(1-\phi)}{N_B}\ln(1-\phi) + \chi\phi(1-\phi). \quad (2)$$

Here $N_A$ and $N_B$ represent the number average polymerization degree of polymer A and emerging polymer B, respectively, and $\chi$ is the temperature dependent interaction parameter.

The dynamics of polymerization-induced phase separation may be described by incorporating the polymerization kinetics into the TDGL equation:[22,28-34]

$$\frac{\partial \phi(\mathbf{r},t)}{\partial t} = \nabla \cdot \Lambda(\phi) \nabla \frac{\delta F}{\delta \phi(\mathbf{r},t)} + \zeta(\mathbf{r},t), \quad (3)$$

where $\zeta$ is the Gaussian noise and satisfies the fluctuation dissipation relation,[30] $\langle\zeta(\mathbf{r},t)\rangle=0$, $\langle\zeta(\mathbf{r},t)\zeta(\mathbf{r}',t)\rangle=-2k_B T\Lambda(\phi)\nabla^2\delta(\mathbf{r}-\mathbf{r}')\delta(t-t')$. $\Lambda(\phi)$ is the mobility and depends on the order parameter.

For a binary system, the mobility is generally given by[35]

$$\Lambda(\phi)=\frac{\Lambda_A\Lambda_B}{\Lambda_A+\Lambda_B}, \tag{4}$$

where $\Lambda_A=\phi N_A D_A$, $\Lambda_B=(1-\phi)N_B D_B$. $D_A$ and $D_B$ are the self-diffusion coefficients of the polymer A and emerging polymer B chains, respectively. Assuming polymer A and emerging polymer B are linear, and neglecting chain entanglement, in the framework of Rouse theory the self-diffusion coefficients are written as[36] $D_i=k_B T/\xi_i N_i$, $i$=A, B. Here $\xi_i$ are the friction coefficients per monomer unit, which for simplicity are assumed to be equal for both species, $\xi_A=\xi_B=\xi_0$. If chain entanglement is considered, in the framework of reptation theory the self-diffusion coefficient is written as[37] $D_i=k_B T N_{e,i}/\xi_i N_i^2$, $N_{e,i}$ being the distances between the entanglements of the components.

Inserting eq.(1) to eq.(3), one obtains[38]

$$\frac{\partial\phi(\mathbf{r},t)}{\partial t}=k_B T\nabla\cdot\Lambda(\phi)\nabla\left(\frac{1+\ln\phi}{N_A}-\frac{1+\ln(1-\phi)}{N_B}+\chi(1-2\phi)-\frac{a^2}{18\phi(1-\phi)}\nabla^2\phi\right)+\zeta. \tag{5}$$

Following Kyu et al.[22] and Rey et al.[25-28], we assume that the growth rate of the reactive component, $N_B$ in eq.(5) can be determined by solving the second reaction kinetics equation

$$\frac{dp(t)}{dt}=k(1-p)^2, \tag{6}$$

where *p* is the extent of reaction and *k* is the reaction rate constant which has the usual Arrhenius form. Eq.(6) describes the self-polycondensation of B type monomer. It should be pointed out that eq.(6) is strictly valid only for a homogenous system. To apply eq.(6) to phase separating system we assume that the extent of reaction is a spatially averaged value. $N_B$ is represented with number average molecular size as

$$N_B = 1/(1-p(t)). \tag{7}$$

Eqs. (5)-(7) describe the governing model for dynamics of PIPS.

When polymerization is initiated in the homogenous state, the mixture undergoes phase separation due to instabilities induced by the increase in extent of reaction of monomer B. Figure 1 shows the resulting phase diagram. The critical point is

$$\phi_c = \sqrt{\frac{1}{2\chi N_A}}, \quad p_c = 1 - 2\chi\left(1 - \sqrt{\frac{1}{2\chi N_A}}\right)^2. \tag{8}$$

It can be seen from figure 1 that for off-critical composition the system will first fall into the metastable region and then enters into the spinodal region. But at the critical composition, the system will enter into spinodal region directly.

Following Kyu et al.[22], we neglect the thermal noise and take account of the concentration fluctuations, viz., $\delta\phi = \phi - \phi_0$. Eventually, the linear Cahn-Hilliard equation, which has been extensively used for the early stage of spinodal decomposition where concentration fluctuations are small, is obtained

$$\frac{\partial \delta\phi}{\partial t} = \Lambda(\phi_0)\nabla^2\left(\frac{1}{N_A\phi_0} + \frac{1}{N_B(t)(1-\phi_0)} - 2\chi - \frac{a^2}{18\phi_0(1-\phi_0)}\nabla^2\right)\delta\phi. \tag{9}$$

From eq.(9), the structure factor $S(q, t)$ is[22]

$$\frac{S(q,t)}{S_q(0)} = \exp\left\{-2\Lambda(\phi_0)q^2\int\left(\frac{1}{N_A\phi_0}+\frac{1}{N_B(t)(1-\phi_0)}-2\chi+\frac{a^2q^2}{18\phi_0(1-\phi_0)}\right)dt\right\}, \quad (10)$$

where $S_q(0)$ is the structure factor at $t=0$. It is reminded that the second derivative of the local free energy, $f'' = \frac{1}{N_A\phi_0}+\frac{1}{N_B(t)(1-\phi_0)}-2\chi = 0$, determines the spinodal curve. $f'' > 0$ is characteristic for the metastable regime between the binodal and spinodal lines, and $f'' < 0$ characterizes the unstable region within the spinodal. We first discuss the case of a jump from the homogeneous state into the spinodal region. This can be realized in a system with a high reaction rate where the passage through the metastable region is fast with respect to the rate of concentration fluctuations. Of course, at the critical composition the system can also directly enter into the spinodal region. In these cases a dominant wave vector $q_m$ (spinodal ring) can be obtained from the maximum of $S(q,t)/S_q(0)$ in eq. (10). By putting $\frac{\partial\left[S(q,t)/S_q(0)\right]}{\partial q} = 0$, one obtains

$$\begin{aligned}q_m^2 &= \frac{9\phi_0(1-\phi_0)}{a^2t}\int\left(2\chi-\frac{1}{N_A\phi_0}-\frac{1}{N_B(t)(1-\phi_0)}\right)dt \\ &= \frac{9\phi_0(1-\phi_0)}{a^2}\left(2\chi-\frac{1}{N_A\phi_0}-\frac{1}{t}\int\frac{dt}{N_B(t)(1-\phi_0)}\right)\end{aligned}. \quad (11)$$

It is clear that a scattering maximum doesn't exist in the metastable region because $f'' > 0$. Unlike thermally induced phase separation, where $q_m$ is constant during the early stage, in PIPS it change with time even in the early stage. If $q_m$ increases with time, we must have $\partial\left(\frac{1}{t}\int\frac{dt}{N_B(t)}\right)\Big/\partial t < 0$, that is to say, we have

$$\frac{t}{N_B(t)}-\int\frac{dt}{N_B(t)} < 0 \quad (12)$$

If eq.(12) is true, the decrease of domain size with time must exist. Let function $g(t) = \frac{t}{N_B(t)} - \int \frac{dt}{N_B(t)}$, the initial condition is $g(0)=0$. $\frac{\partial g(t)}{\partial t} = -\frac{t}{N_B(t)^2} \frac{\partial N_B(t)}{\partial t} < 0$ because of the increase of $N_B$ with time ($\frac{\partial N_B}{\partial t} > 0$), then function $g(t)$ decreases with $t$, that is to say, eq.(12) is always true, independent of the specific reaction kinetics. Thus the reduction of initial length scale is due to the increase of $N_B$ for a fast passage through the metastable region or entering into the spinodal region directly through the critical point. It also indicates that NISD is not necessarily the only mechanism for the reduction of initial length scale.

The case that the passage through the metastable region occurs at finite reaction rates may be qualitatively discussed as follows. In this case $f'' > 0$, and fluctuations of all wave vectors decay. As can be seen from eq.(10) the decay rate decreases with increasing $N_B$ or reaction degree $p$. Some of the modes corresponding to individual p-values situated in the metastable region may have survived when the system is passing through the spinodal line (see Fig. 1) and they may couple them to a new stable modes being created in the spinodal region. The extent to which modes survive at the time of entrance into the spinodal region depends on the residence time of the system in the interval between the binodal and the spinodal, i. e. on the reaction rate. If the reaction is very slow, most of the modes will have decayed, while for very fast reactions the amplitudes will also be small because they have no time to build up. However, for intermediate reaction rates a coupling of modes created in the metastable region to spinodal decompostion cannot be excluded. In order to discuss the situation of finite reaction rates we have performed simulations using the nonlinear TDGL equation (5).

**III. Simulation**

In our simulation, we start with the initially uniform phase, the thermal noises in eq. (5) is necessary to avoid the decay-out of fluctuations before the polymerization-induced phase separation initiates. We introduce thermal fluctuation through a conserved noise term in the current.[39] In addition, eq.(5) and eq.(7) are scaled in a dimensionless form with $a_0=a$ and $\tau_0=a_0^2/(\Lambda_0 k_B T)$ as length and time scale respectively, where $\Lambda_0=\xi_0/k_B T$ is a scaling constant and has units of mobility. Eventually, the dimensionless equation is

$$\frac{\partial \phi(\mathbf{r},t)}{\partial t} = \nabla \cdot \frac{\Lambda(\phi)}{\Lambda_0} \nabla \left( \frac{1+\ln\phi}{N_A} - \frac{1+\ln(1-\phi)}{N_B} + \chi(1-2\phi) - \frac{1}{18\phi(1-\phi)} \nabla^2 \phi \right) + \nabla \cdot \eta , \quad (14)$$

with $\langle \eta(\mathbf{r},t) \rangle = 0$ and $\langle \eta(\mathbf{r},t) \eta(\mathbf{r}',t') \rangle = -\varepsilon \frac{\Lambda(\phi)}{\Lambda_0} \nabla^2 \delta(\mathbf{r}-\mathbf{r}') \delta(t-t')$,[39] where $\varepsilon$ is the magnitude of the fluctuation. In this paper, $\varepsilon$ is set as 0.0001.

The eq.(14) is solved numerically on a two dimensional $128 \times 128$ grid. For a spatial step, a standard central difference discretization scheme was used. An explicit method was utilized for a time step. Both the grid size and time step were chosen sufficiently small to ensure that changes in them exert no effect on the simulation results. A periodic boundary condition was imposed in both spatial directions. In the simulation, the spatial step is set as $\Delta x = 1.0$, the time step $\Delta t = 0.01$. The initial condition for the order parameter field $\phi(\mathbf{r},0)$ in each run consisted of uniformly distributed small amplitude fluctuations about an average value $\phi_0$. Thermal noise was mimicked by uniformly distributed random numbers with amplitude $\sqrt{3\varepsilon/\Delta x^d \Delta t}$. The other parameters are set as $N_A$=100, $\chi$=0.2, $D_A$=4$k_BT$/$\xi_0 N_A$, $D_B$=4$k_BT$/$\xi_0 N_B$. According to eq.(8), the critical composition is $\phi_c$=0.1581.

In order to trace the domain growth, the average size of domains $R(t)$ is calculated by[40-42]

$$R(t) = \left( \frac{\int d\mathbf{q}\, q^2 S(\mathbf{q},t)}{\int d\mathbf{q}\, S(\mathbf{q},t)} \right)^{-1/2}, \tag{13}$$

where $S(\mathbf{q},t) = \langle \phi_\mathbf{q}(t)\phi_{-\mathbf{q}}(t) \rangle$ is the structure factor at time $t$.

First, we consider different off-critical compositions with the reaction rate constant $k$=0.01. As polymerization advances, the system will first pass through metastable region and then drift to the spinodal region. Figure 2 shows the typical time evolution of the simulated pattern for initial volume fraction of polymer A $\phi_0$= 0.3. For quantification, it is customary to investigate the scattering patterns. Figure 3 is the typical snapshots of the scattering pattern for the morphological patterns shown in Figure 2. The initial scattering pattern is small and very diffuse without a clear maximum (see $t$=1000). The diameter of the scattering pattern increases with progressive polymerization up to $t$=3000, which reflects the decrease of the average domain size and is in agreement with the morphological pattern in real space. As the polymerization continues, the intensity of the scattering ring increases while the size collapses to a smaller diameter due to the structure coarsening ($t$>3000). The details of growth dynamics and morphological evolution can be more quantitatively characterized by the average domain size $R(t)$ defined by Eq. (13), which is shown as a function of time on a log-log plot in Figure 4 for $\phi_0$=0.3, 0.55 and 0.7, respectively. From Figure 4, it is clearly seen that for all compositions, $R(t)$ rises rapidly first, then decreases in the early stage of the phase separation, followed by domain growth in the late stage of the phase separation. Here we should point out that Rey et al.[29] also observed the increase of the $q_m$ in the early stage of PIPS in their simulation, which means the reduction of the initial length scale. From the theoretical analysis and simulation, we confirm that the initial length scale will decrease with the extent of reaction, which is also in agreement with experimental results.[16,22]

Second, another important parameter that can influence the dynamics and morphology of phase separation is the reaction rate constant. Here we consider the case $\phi_0$=0.3 with different reaction rate constants $k$=0.001, 0.01 and 0.1, respectively, as shown in Figure 5. It is seen that the induction time decreases with the reaction time and the decrease of domain size in the early stage of phase separation is also observed for different reaction rate constants. Finally, the simulation is carried out at critical composition $\phi_c$. At the critical point, with increasing the extent of reaction, the system is directly thrown into unstable region, without passing through metastable region. Figure 6 shows the evolution of the average domain size $R(t)$. It is interesting that the reduction of initial length scale is also observed, which agrees with our theoretical results. In the late stage, the phase separation is suppressed which agrees with previous prediction[27].

All above simulation results (Figures 4-6) show the similar behavior of the domain size as a function of the time. Physically, the competition of two opposite effects determines the observed behavior. On the one hand, phase separation requires the coarsening of domains. On the other hand, the chemical quench depth increases with increasing $N_B$, which suppresses the domain coarsening. At the early stage of the phase separation, the second effect is dominant which results in the reduction of the initial length scale. From the temporal evolution of morphological patters, it is due to the large irregular shape regions forming the interconnected structure pinching to form droplets. Finally, the coarsening of droplets is more important, resulting in the increase of the domain size. Here, we should point out that even if the system locates initially in spinodal region, the reduction of initial length scale also occurs. This further indicates that the NISD is not the only mechanism for the reduction of initial length scale. Thus it is not correct to consider the reduction of initial length scale as the proof that the phase separation goes through nucleation and growth mechanism.

**IV. Conclusion**

In this paper, we have analytically and numerically investigated the dynamics and the morphology of polymerization-induced phase separation in the system consisting of polymers and reactive monomers. By incorporating the reaction kinetics into the time-dependent Ginzburg Landau equation, we carried out the theoretical analysis and simulation on a two-dimensional square lattice. Both theoretical and numerical results show the existence of the reduction of initial length scale for the early stage of phase separation, which is independent of the fact whether the system goes through the metastable region or not. Thus nucleation and growth in the metastable region precedent to the spinodal decomposition in the unstable region is not prerequisite for the reduction of initial length scale in PIPS.

**ACKNOWLEDGMENTS** The author thanks W. Gronski, C. Friedrich, P. Arends and Y. Yang for fruitful discussions. The work received financial support by the SFB 428 of the "Deutsche Forschungsgemeinschaft".

**Figure Captions:**

**Figure 1.** Binodal and Spinodal curves in extent of reaction vs composition transformation diagram for $N_A$=100 and $\chi$=0.2. The critical point and the location of stable, metastable and unstable regions are shown.

**Figure 2.** Temporal evolution of morphological patterns with reaction rate constant $k$=0.01 for initial volume fraction of polymer A $\phi_0$=0.3. Polymer A rich regions are shown bright, and emerging polymer B rich regime are shown dark.

**Figure 3.** Temporal evolution of scattering patterns for Figure 2.

**Figure 4.** Temporal evolution of the average domain size with reaction rate constant $k$=0.01 for off-critical compositions.

**Figure 5.** Temporal evolution of the average domain size with different reaction rate constants $k$ for off-critical composition $\phi_0$=0.3.

**Figure 6.** Temporal evolution of the average domain size with reaction rate constant $k$=0.01 for critical composition, $\phi_0$=0.1581.

**References:**

1) J. D. Gunton, M. S. Miguel, and P. S. Sahni, "*Phase transition and critical Phenomena*", edited by C. Domb and J. H. Lebowitz (Academic, London, 1983) Vol. 8.

2) A. J. Bray, *Adv. Phys.* **43**, 357 (1994).

3) P. C. Hohenberg and B. I. Halperin, *Rev. Mod. Phys.* **49**, 435 (1977).

4) A. Onuki, *J. Phys. Condens. Matter* **9**, 6119 (1997).

5) I. M. Lifshitz and V. V. Slyozov, *J. Phys. Chem. Solids* **19**, 35 (1961).

6) E. D. Siggia, *Phys. Rev. A* **20**, 595 (1979).

7) S. C. Glotzer, D. Stauffer, N. Jan, *Phys. Rev. Lett.* **72**, 4109 (1994).

8) S. C. Glotzer, E. A. Di Marzio, M. Muthukumar, *Phys. Rev. Lett.* **74**, 2034.(1995)

9) Q. Tran-Cong and A. Harada, *Phys. Rev. Lett.* **76**, 1162 (1996).

10) A. Harada and Q. Tran-Cong, *Macromolecules* **30**, 1643 (1997).

11) Q. Tran-Cong, T. Ohta, and O. Urakawa, *Phys. Rev. E* **56**, R59 (1997).

12) Q. Tran-Cong, J. Kawai, Y. Nishikawa, and H. Jinnai, *Phys. Rev. E* **60**, R1150 (1999).

13) R. J. J. Williams, B. A. Rozenberg, J. P. Pascault, *Adv. Polym. Sci.*, **128**, 95 (1997).

14) T. Inoue, *Prog. Polym. Sci.* **20**, 119 (1995).

15) J. Zhang, H. Zhang, Y. Yang, *J. Appl. Polym. Sci.* **72**, 59 (1999).

16) L. Q. Peng, J. Cui, S. J. Li, *Macromol. Chem. Phys.* 201, 699 (2000)

17) L. H. Sperling, "*Polymeric Multicomponent Materials; An Introduction*" (John Wiley & Sons, New York 1997)

18) J. Y. Kim, C. H. Cho, P. Palffy-Muhoray, M. Mustafa, and T. Kyu, *Phys.Rev. Lett.* **71**, 2232 (1993).

19) M. Okada, K. Fujimoto and T. Nose, *Macromolecules* **28**, 1795 (1995).

20) M. Okada and T. Sakaguchi, *Macromolecules* **32**, 4154 (1999).

21) M. Okada and T. Sakaguchi, *Macromolecules* **34**, 4027 (2001).

22) T. Kyu, J. H. Lee, *Phys. Rev. Lett.* **76**, 3746 (1996).

23) Y. M. Zhu, *Phys. Rev. E* **54**, 1645 (1996).

24) X. Y: Wang, Y. K. Yu and P. L. Taylor, *J. Appl. Phys*. **80**, 3285 (1996)

25) J. C. Lin and P. L. Taylor, *Phys. Rev. E* **49**, 2476 (1994).

26) J. C. Lin and P. L. Taylor, *Mol. Cryst. Liq. Cryst.*, **237,** 25 (1993).

27) T. Ohnaga, W. Chen, T. Inoue, *Polymer*, **35**, 3774 (1994).

28) P. K. Chan, A. D. Rey, *Macromolecules* **29**, 8934 (**1996)**.

29) P. K. Chan, A. D. Rey, *Macromolecules* **30**, 2135 (1997).

30) J. Oh, A. D. Rey, *Macromol. Theory Simul.* **9**, 641 (2000).

31) M. Okada, H. Masunaga, and H. Furukawa, *Macromolecules* **33**, 7238 (2000).

32) T. Kyu and H. W. Chiu, *Polymer* **42**, 9173 (2001).

33) D. Nwabunma, H.W. Chiu, and T. Kyu, *J. Chem. Phys*. 113, 6429 (2000)

34) T. Kyu, D. Nwabunma, and H. W. Chiu, *Phys. Rev. E* **63**, 061802 (2001).

35) M. Takenaka and T. Hashimoto, *Phys. Rev. E* **48**, R647 (1993).

36) P. E. Rouse, *J. Chem. Phy*s. **21**, 1272 (1953).

37) P. G. de Gennes, *J. Chem. Phys*. **72**, 4756 (1980).

38) K. Binder, J. Chem. Phys. 79, 6387 (1983)

39) J. Sharma and S. Puri, *Phys. Rev. E* **64**, 021513 (2001).

40) F. Corberi, G. Gonnella and A. Lamura, *Phys. Rev. Lett.* **83**, 4057 (1999); *ibid.*, **81**, 3852 (1998)

41) K. F. Luo, Y. L. Yang, *J. Chem. Phys*. **115**, 2818 (2001)

42) K. F. Luo, Y. L. Yang, *Macromolecules* **35**, 3722 (2002)

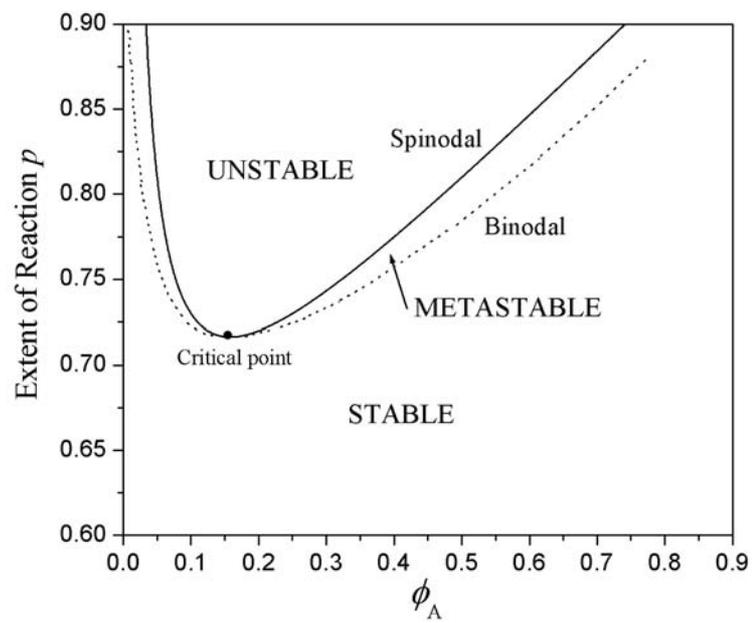

Figure 1.

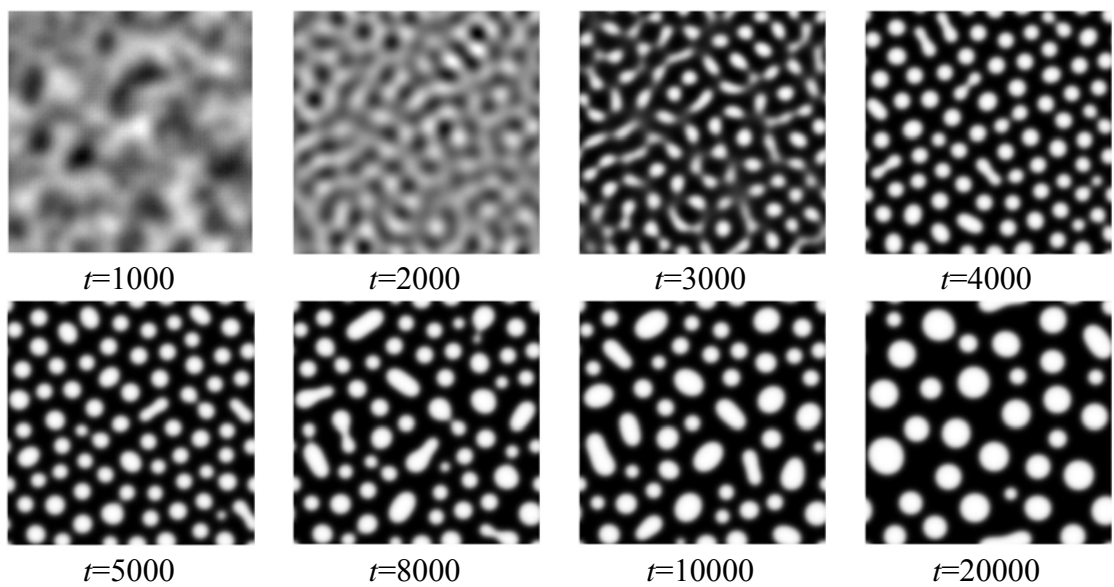

Figure 2.

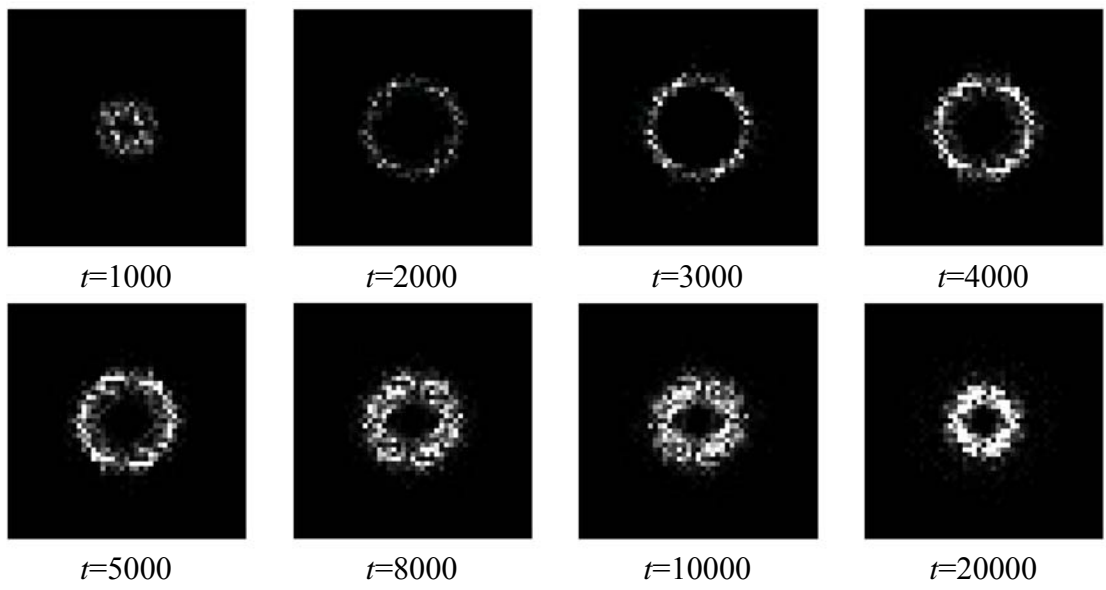

| | | | |
|---|---|---|---|
| $t=1000$ | $t=2000$ | $t=3000$ | $t=4000$ |
| $t=5000$ | $t=8000$ | $t=10000$ | $t=20000$ |

Figure 3.

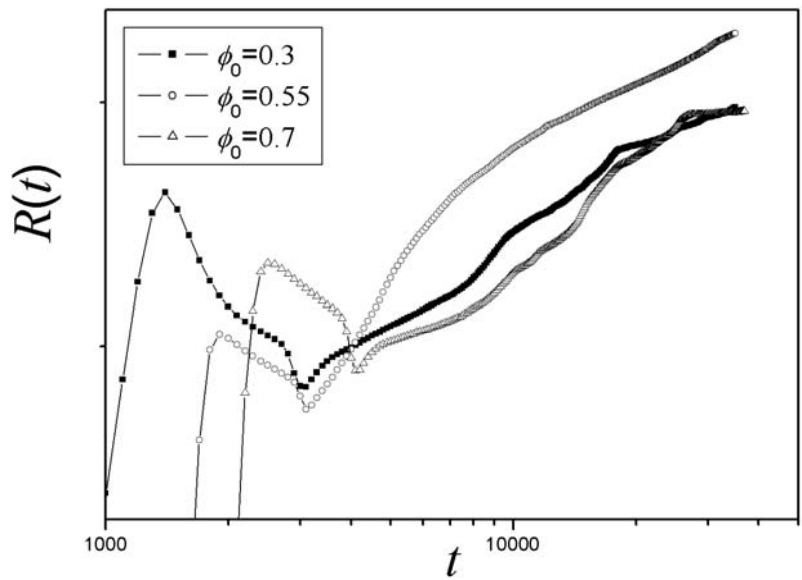

Figure 4.

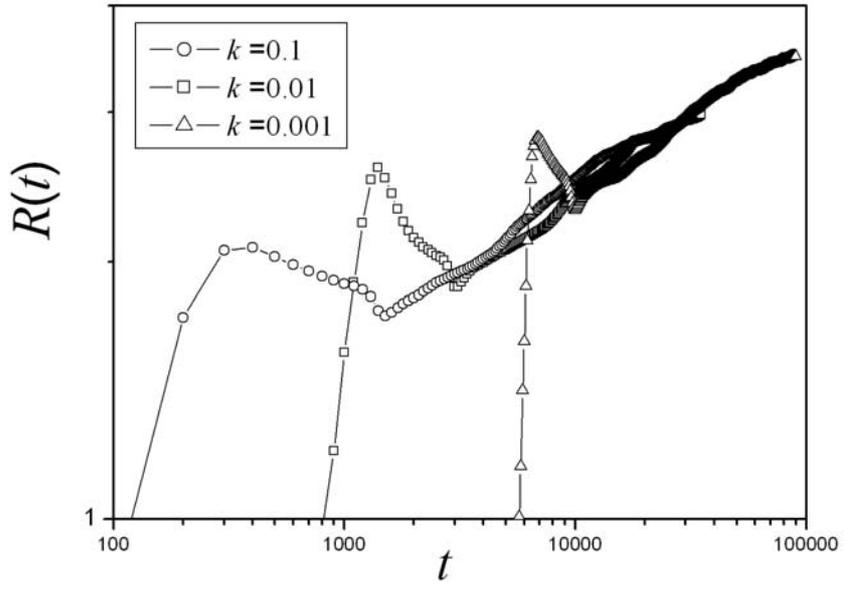

Figure 5.

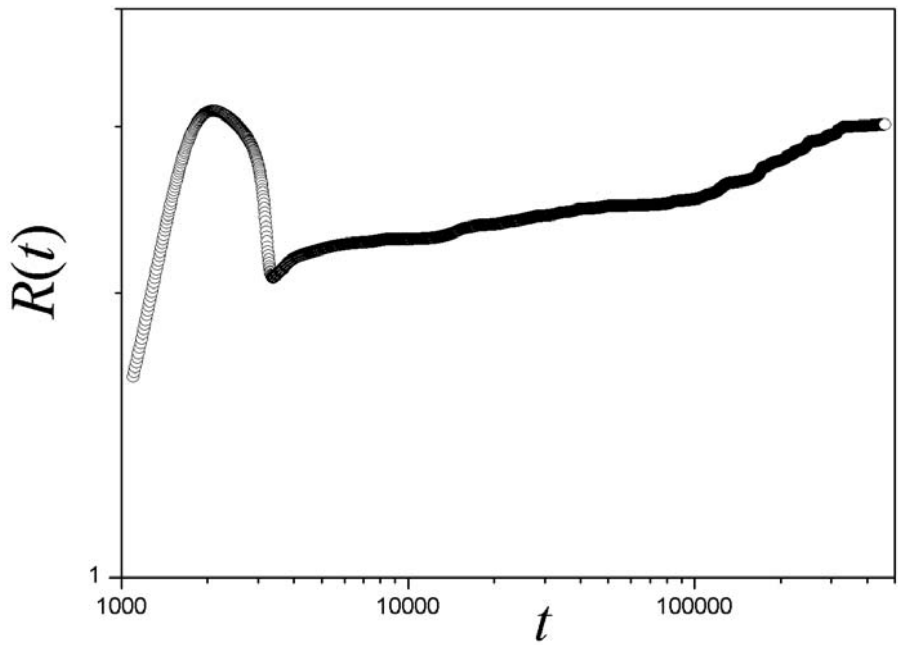

Figure 6.